\documentclass[amsfonts,11pt]{article}
\title{Linearized Quantum Gravity Using \\the Projection Operator Formalism}
\author{W. R. Bomstad and John R. Klauder$\footnote{Also, Department of Mathematics.}$\\
Department of Physics\\
University of Florida\\
Gainesville, FL  32611}

\usepackage{amsmath,amssymb}

\newcommand{\me}{\mathrm{e}}

\date{}
\begin{document}
\maketitle
\numberwithin{equation}{section}

\begin{abstract}
The theory of canonical linearized gravity is quantized using the
Projection Operator formalism, in which no gauge or coordinate
choices are made.  The ADM Hamiltonian is used and the canonical
variables and constraints are expanded around  a flat background. As
a result of the coordinate independence and linear truncation of the
perturbation series, the constraint algebra
 surprisingly becomes partially second-class in both the classical and quantum pictures after all secondary constraints are
 considered.
While new features emerge in the constraint structure, the end
result is the same as previously reported: the (separable) physical
Hilbert space still only depends on the transverse-traceless degrees
of freedom.
\end{abstract}

\section{Introduction}

In the early days of canonical quantum gravity (CQG), it was widely
thought that the advent of a consistent ADM-Hamiltonian description
of general relativity would herald the successful merger between
quantum mechanics and general relativity, subsequently providing a
useful description of Planck-scale physics.  Instead of the
Schr\"{o}dinger equation, CQG had the Wheeler-DeWitt equation (WdW)
$\cite{DeW}$, widely regarded as the functional analogue of the
Schr\"{o}dinger equation. However, the promise of solutions to the
Wheeler-DeWitt equation has been indefinitely postponed due to the
many illnesses plaguing the procedure.  The diseases of
non-renormalizability, constraint consistency, and the problem of
time have caused many physicists to abandon the thought of canonical
quantum gravity altogether; or, at the very least, formulate CQG in
terms of a completely different set of variables.

One of these persistent problems has been the problem of constraint
quantization.  How can one consistently promote the constraints of
general relativity to self-adjoint quantum operators, and at the
same time be certain these constraints are still satisfied by the
quantum dynamics? An entire constraint classification system was
invented by Dirac to address this very issue. In his exploration
$\cite{dirac}$, it was noticed that, while the constraints of
gravity are closed under the classical Poisson algebra, when the
constraints are promoted to quantum operators, a subset of the
constraints mutated into a different class of constraints
altogether.  Or, in the parlance of Dirac, the {\it first class}
constraint functions transmuted into {\it second class} constraint
operators ( see, e.g., $\cite{gov}$ or $\cite{teithenn}$ for a
modern treatment). For a time, it was believed that with a
consistent factor-ordering of the Hamiltonian constraint of general
relativity, that constraint consistency would be maintained
$\cite{wood}$.

To the present day,
 the CQG program, even in its modern versions $\cite{ash}$, has encountered numerous difficulties associated with
 problem of constraint quantization, the anomalous constraint algebra being only one such problem.  Another
problematic aspect of quantizing gravitational constraints is that
they are infinite in number, as each constraint is needed per point
in space. Any modern canonical quantum gravity program using metric
variables must therefore be equipped to quantize constraints which
are both infinite in number and partially second class in their
commutator algebra.

The Projection Operator method of quantization provides the
machinery in which the Affine Quantum Gravity (AQG) program, a
variation of the CQG Program, proposes to quantize the constraints
of general relativity. Prior work has demonstrated that, for a
multitude of finite dimensional toy models $\cite{JRKearly, JRKQCS}$
including one which mimics the anomalous constraint algebra of
gravity $\cite{lilklauder}$, the Projection Operator provides an
unambiguous reduction from the unconstrained Hilbert space to the
physical Hilbert space. It suffices to say that much of the success
involved in using the Projection Operator to quantize these systems
comes from the ability to quantize first and second class systems of
constraints within one general formulation.  The key to this unified
treatment of constraints is that no gauge choices are to be made at
the classical level.  One must quantize the theory before any
systematic reduction takes place.

 The initial goal of this work was to further examine the feasibility of quantizing
 an infinite number of constraints with the Projection Operator quantization
 formalism.
Linearized general relativity, a theory whose quantized form is well
known, was selected as the ideal candidate for testing the
techniques that the Projection Operator formalism proposes for the
full, nonlinear theory of gravity.  However, the stipulation of
gauge independence at the classical level necessitated the inclusion
of secondary constraints into the theory. These new constraints then
led to a second class constraint sector in both the classical and
quantum theories. The final result of this work shows that the
Projection Operator can quantize infinite numbers of (non-anomalous)
first and second class constraints in a quantum theory.

 The next section briefly reviews the classical
theory of linearized gravity, and highlights how
coordinate-independence changes the constraint structure.  The third
section quantizes the system following the Projection Operator
formalism, discussing salient features of reproducing kernels,
Projection Operators and coherent states along the way.  The last
section of the text concludes and compares our work to some past
work done in the CQG program. Finally, there are three appendices: one
appendix which provides a simple toy model analogue of the analysis
required in the paper, another which proves a theorem used to show the reduction of reproducing kernels of quantum, first-class constraints, and a third appendix which lists the
fundamental algebras used in the main text.

\section{Classical theory}

This section consists of a review of linearized gravity, the
perturbation of the metric field around a flat background; see $\cite{ADM}$ or
$\cite{wald}$.  We
follow the prescription of quantizing the theory $\emph{before}$
reducing the space. Hence, no gauge is chosen, and the flat,
canonical phase space $\Gamma$ is not mapped to a reduced phase
space $\Gamma_R$. The physical degrees of freedom will only become
apparent in the quantum world after the Hilbert space $\mathfrak{H}$
is reduced to the physical Hilbert space $\mathfrak{H}_P$ with the
help of a suitable projection operator.
  There is a good reason for quantization before reduction since there are counterexamples which show that reduction and
quantization do not necessarily commute \cite{JRKQCS}.  In all work to follow, surface terms will be discarded.

\subsection{Geometrodynamics}

In order to craft general relativity into a Hamiltonian formalism,
it is well known that the manifest covariance of the
Einstein-Hilbert action can be (3+1)-decomposed and encoded into the
ADM $\cite{ADM}$ action.  The resulting theory
 describes the evolution of three-dimensional
hypersurfaces embedded in four-dimensional spacetime. The dynamical
variables are the symmetric 3-metric of the hypersurface and its
canonical momentum density, meaning that the phase space $\Gamma$
for this theory is $\mathbb{R}^{12}$ at each spatial point. The
initial data of a metric, the symmetric tensor
$g_{ab}(x)\footnote{In each of these dynamical variables, and those
that follow, lower case Latin indices indicate Euclidean, spatial indices (e. g.
$i\in \{1,2,3\}$), and repeated indices in a term are summed over.
Greek indices, such as $\mu \in \{0,i\}$, are of the Minkowski, spacetime variety. }$, with its canonical
momentum density tensor $\pi^{ab}(x)$, are specified and a set of
constraints are satisfied by this data. The evolution of the
hypersurface is thereafter restricted to being causal and invariant
to diffeomorphisms on the hypersurface, a spacelike 3-surface.

The ADM action is defined as
\begin{eqnarray}
I[\pi,g]= \int dt \;\int d^3x \;\left[ \pi^{ab}\dot{g}_{ab} -N^i H_i
-N H
  \right],
\label{ADMAction}
\end{eqnarray}
where the lapse $N=N(x)$ and shift vector $N^i=N^i(x)$ are the
Lagrange multipliers.  The constraints are then given by $H(x)=0$
and $H_i(x)=0$, and are defined respectively as the Hamiltonian
constraint and the set of diffeomorphism constraints. With
$\pi\equiv \pi^a_{\;a}$, the Hamiltonian constraint term is given by
\begin{eqnarray}
 C[N]&\equiv& \int d^3x\;N(x)\:H (x)\label{CN}\\&=& \int d^3x \;N\left\{- \sqrt{g} \left[ R+ \frac{1}{g} \left(
\frac{\pi^2}{2} - \pi^{ab}
  \pi_{ab} \right) \right]\right\} \nonumber
\end{eqnarray}
where $R$ is the three-dimensional Ricci scalar and follows the same
convention of index contraction as \cite{MTW}. Equation (\ref{CN})
provides a fundamental link between the intrinsic and extrinsic
curvature of the evolving hypersurfaces, one that must remain
constant throughout time \cite{kook2}. Infinitesimal diffeomorphisms
in the initial data are generated by the diffeomorphism constraint
term
\begin{eqnarray}
C[N^a] &\equiv &\int d^3x \;N^a(x)\:H_a(x)= \int d^3x\;\left[-2N^a\pi_a^{\;\;\;k} \vert_k \right] \label{H_i}\\
&=& \int d^3x \;N^a\left[-2 \pi^{\;\;k}_{a\;\;\;,k} - (2g_{al,m}-
g_{lm,a})\pi^{lm} \right] \nonumber .
\end{eqnarray}
These diffeomorphisms are the generators of small tangential
displacements on the spacelike hypersurfaces.

\subsection{Perturbation of metric variables}

The metric tensor, and its conjugate momenta can be expanded around
a flat background according to
\begin{eqnarray}
g_{ab} (x) &\to& \delta_{ab} + \epsilon \; h^{(1)}_{ab}(x) + \epsilon^2h^{(2)}_{ab}(x) + O(\epsilon^3),\nonumber \\
\pi_{ab} (x) & \to& 0 + \epsilon \; p^{(1)}_{ab}(x)  + \epsilon^2
p^{(2)}_{ab}(x) +O(\epsilon^3), \label{Gpert}
\end{eqnarray}
where $\epsilon$ is merely an order parameter for the perturbation
analysis, and $\delta_{ab}$ is the Euclidean 3-metric. The lapse and
shift are similarly expanded according to
\begin{eqnarray}
N(x)&\to& 1+ \epsilon N^{(1)}(x)+ \epsilon^2N^{(2)}(x) + O(\epsilon^3),\nonumber \\
N_a(x) &\to& 0 + \epsilon N_a^{(1)}(x) + \epsilon^2 N_a^{(2)}(x) +
O(\epsilon^3) ,\label{Npert}
\end{eqnarray}
c.f., \cite{kook}.

Implementing an orthogonal decomposition for the symmetric tensors
into their transverse-traceless,  transverse, and longitudinal
components provides an elegant, reduced expression for these
constraints. Each $p_{ab}$ and $h_{ab}$ tensor is decomposed into a
set of orthogonal components $\cite{ADM}$ represented by
\begin{equation}
f_{ab}(x)= {\bf f}_{ab}^{TT}(x)+ {\sf f}_{ab}^T(x) +
2\mathfrak{f}_{(a,b)}^L(x), \label{tensordcomp}
\end{equation}
where $f_{ab}(x)$ stands for either the metric or momentum density tensor.
Also note the use of different fonts to denote the various tensor
components; as we proceed, we shall continue to use the different
fonts and drop the capital superscripts.

In ($\ref{tensordcomp}$), each set of components has the same
definition as in \cite{ADM}.  The components of which are uniquely
determined by (\ref{tensordcomp}) and the relations
\begin{eqnarray}
 {\bf f}_{ab,b} &\equiv& 0, \; \quad \;
{\bf f}_{aa}\equiv 0 \label{td1}\\
{\sf f}_{ab}&=&\frac{1}{2} \left( {\sf f}\delta_{ab}- \frac{\partial_a \partial_b}{\nabla^2} {\sf f}\right) \label{td2}\\
\mathfrak{f}_a&=&\frac{1}{\nabla^2}\left( f_{ab,b} - \frac{f_{bc,bca}}{2\nabla^2}\right), \label{td3}
\end{eqnarray}
yielding two degrees of freedom in ${\bf f}_{ab}$, one degree of
freedom in ${\sf f}={\sf f}_{aa}$, and three in the vector
$\mathfrak{f}_a$. This type of symmetric tensor decomposition allows
the linear constraints to be written in a more streamlined fashion.

Using such a decomposition along with the expansion strategy given
in ($\ref{Gpert}$) and ($\ref{Npert}$), the integrand in
(\ref{CN}) may be written to second order as
\begin{eqnarray}
N(x)\;H(x)&\to& -\nabla^2  {\sf h}^{(2)}+ R^{(1)} +h_{aa}\: R^{(1)}/2 +
R^{(2)} \\ &\;& + p_{ab}p^{ab}-p^2/2-N^{(1)}\nabla^2{\sf h}^{(1)}
\label{NH}.
\end{eqnarray}
The factor $N^{(1)}(x)$ is to be interpreted as one of the new Lagrange multipliers
of the linearized theory. The linear Hamiltonian constraint density
can then be immediately read off as the term multiplying
$N^{(1)}(x)$, namely,
\begin{eqnarray}
H^{(1)}(x)=-\nabla^2 {\sf h}^{(1)}(x).  \label{H0lin}
\end{eqnarray}
The rest of the terms in (\ref{NH}) can then be interpreted as the negative of an unconstrained Hamiltonian density, defined by
\begin{eqnarray}
\mathcal{H}[p,h]= - R^{(1)} -h_{aa}\: R^{(1)}/2 - R^{(2)} - p_{ab}p^{ab}+p^2/2. \label{curlyH}
\end{eqnarray}
Using the decomposition of (\ref{tensordcomp})-(\ref{td3}), it has
been shown that the only transverse-traceless terms in
(\ref{curlyH}) combine to form a harmonic oscillator piece
\cite{kook}:
\begin{equation}
\mathcal{H}_{TT}[{\bf p ,h}]={\bf p}^{(1)}_{ab}{\bf
p}^{(1)}_{ab}+\frac{1}{4}{\bf h}^{(1)}_{ab,c}{\bf h}^{(1)}_{ab,c}
.\label{kamel}
\end{equation}
However, if no gauge is fixed, unconstrained Hamiltonians for the
transverse and longitudinal variables arise from $\mathcal{H}$
 and must also be considered in the dynamics.

Temporarily postponing the discussion about the remaining terms in
$\mathcal{H}$, the symmetric tensor decomposition and expansions are
repeated on the term in the action containing the diffeomorphism
constraint. This constraint depends only on the longitudinal portion
of $p_{ab}$ and $h_{ab}$, denoted by the 3-vectors $\mathfrak{p}_a$
and $\mathfrak{h}_a$.  These three degrees of freedom may be
isolated further by splitting each longitudinal vector into
transverse ($\mathfrak{p}^t_a,\mathfrak{h}^t_a$) and longitudinal
pieces ($\mathfrak{p}^l,\mathfrak{h}^l$). With this secondary,
vector decomposition, the expansion of the integrand of
($\ref{H_i}$) becomes
\begin{eqnarray}
N^a(x)\;H_a(x)\to -2 N^{a\;(1)}(x)
\nabla^2\left(\mathfrak{p}^{(1)t}_a +
\mathfrak{p}^{(1)l}_{,a}\right)
\end{eqnarray}
through quadratic order in the expansion.  The linearized
diffeomorphism constraint is therefore
\begin{equation}
H^{(1)}_a(x)=-2\nabla^2\left(\mathfrak{p}^{(1)t}_a(x)+\mathfrak{p}^{(1)l}_{,a}(x)\right)=2\nabla^2\mathfrak{p}_a(x).\label{Hilin}
\end{equation}
Since all the terms in the ADM action have now been expanded through
quadratic order, and only the first order terms contributed, the
superscript $^{(1)}$ will be discarded to allow for shorter
expressions.  A step that can be taken to further simplify the analysis is to work
in momentum space with the Fourier transform pair
\begin{eqnarray}
f_{ab}(x)&=&\int \frac{d^3k}{(2\pi)^{3/2}} \; \widetilde{f}_{ab}(t,{\bf k}) \me^{i {\bf k}\cdot {\bf x}}\\
\widetilde{f}_{ab}(t,{\bf k})&=& \int \frac{d^3x}{(2\pi)^{3/2}}
 f_{ab}(x)\me^{-i {\bf k}\cdot {\bf x}}.
\end{eqnarray}
In momentum space, the constraints of (\ref{H0lin}) and (\ref{Hilin}) transparently give
\begin{eqnarray}
\widetilde{{\sf h}} (t,{\bf k}) =0 = \widetilde{\mathfrak{p}}_a (t,{\bf k}).
\end{eqnarray}

The next step is to find out to which Dirac constraint class the
constraints (\ref{Hilin}) and (\ref{Hilin}) belong.  This algebra
depends on the equal-time canonical Poisson bracket expression
\begin{equation}
\lbrace
h_{ab}(x),p^{cd}(x')\rbrace=\delta^c_{(a}\delta^d_{b)}\delta(x-x').
\label{brack}
\end{equation}
To check that the time derivatives of the constraints are zero, the
Dirac check on constraint consistency, we first must know how to
compute the Poisson brackets of the tensor components. While using
the tensor decomposition of (\ref{tensordcomp})-(\ref{td3}) will
give the same results, we choose to work with
 a more calculationally friendly form in {\bf k}-space.  The complex, null vector $m_a$, its conjugate $\bar{m}_a$,
 and the longitudinally pointing $k_a$ form a basis in {\bf k}-space, such that
\begin{eqnarray}
m_a \:m_a=0=\bar{m}_a\:\bar{m}_a= k_a\: m_a= k_a \bar{m}_a, \quad
m_a \: \bar{m}_a=1.
\end{eqnarray}
This allows for the expansion of each Fourier-transformed canonical variable into its basic tensor components according to
\begin{eqnarray}
\widetilde{f}_{ab}(t,{\bf k})= \widetilde{{\bf f}}_{ab}
+m_{(a}\bar{m}_{b)}\widetilde{{\sf f}} -2i \left(
k_{(a}m_{b)}\widetilde{\mathfrak{f}}^{t\:}_1 + k_{(a}\bar{m}_{b)}\widetilde{\mathfrak{f}}^{t\:}_2
 -i k_{(a}k_{b)}\widetilde{\mathfrak{f}}^{l\;}\right). \label{kdecomp}
\end{eqnarray}
Contractions of combinations of $k_a, m_a ,\bar{m}_a$ into
(\ref{brack}) lead to the algebra in Appendix C. Using the
decomposition of (\ref{kdecomp}) allows the linearized constraints
of (\ref{H0lin}) and (\ref{Hilin}) to be respectively written in
{\bf k}-space as
\begin{eqnarray}
\widetilde{H}(t,{\bf k})&=&  k^2 \widetilde{{\sf h}}(t,{\bf k }) \label{kH}\\
\widetilde{H}_a(t,{\bf k})&=& 2 k^2\left(
\widetilde{\mathfrak{p}}^{\:t}_a(t,{\bf k }) -ik_a
\widetilde{\mathfrak{p}}^{\; l}(t,{\bf k })\right).\label{kHi}
\end{eqnarray}

Lastly, the expansion (\ref{kdecomp}) allows the Fourier transform
of the Hamiltonian density of (\ref{curlyH}) to be written as
\begin{equation}
\widetilde{\mathcal{H}}(t,{\bf k})=
\widetilde{\mathcal{H}}_{TT}+\widetilde{\mathcal{H}}_{T}+
\widetilde{\mathcal{H}}_{L}+\widetilde{\mathcal{H}}_{int}.
\label{Hunc}
\end{equation}
The $\mathcal{H}_T$ and $\mathcal{H}_L$ terms are given by
\begin{eqnarray}
\widetilde{\mathcal{H}}_T(t,{\bf k})&=& |\widetilde{{\sf p}}|^2/2 \label{HT}\\
\widetilde{\mathcal{H}}_L(t,{\bf k})&=&
4k^2|\widetilde{\mathfrak{p}}^{\:t}_a|^2 \label{HL} + 2k^4
|\widetilde{\mathfrak{p}}^{\: l}|^2 -k^4
|\widetilde{\mathfrak{h}}_a|^2 ,
\end{eqnarray}
while $\widetilde{\mathcal{H}}_{TT}$ is the transform of
(\ref{kamel}), or
\begin{equation}
\widetilde{\mathcal{H}}_{TT}(t,{\bf k})= \bar{\widetilde{{\bf
p}}}_{ab}\widetilde{{\bf p}}_{ab} + \frac{k^2}{4}
\bar{\widetilde{{\bf h}}}_{ab}\widetilde{{\bf h}}_{ab}. \label{HTT}
\end{equation}
Finally, the interaction Hamiltonian density is named as such
because it contains interaction amongst the constrained variables in
the form of
\begin{equation}
\widetilde{\mathcal{H}}_{int}(t,{\bf k})=k^2 \left(
\bar{\widetilde{\mathfrak{p}}}^{\;l} \widetilde{{\sf p}} +
\widetilde{\mathfrak{p}}^{\;l}\: \bar{\widetilde{{\sf p}}}\right).
\end{equation}

Next the time derivatives of the constraints are calculated in a
straightforward manner using (\ref{kdecomp}), (\ref{Hunc}), and the
algebra in Appendix C. For the system to continuously remain on the
constraint hypersurface as time evolves, all time derivatives of the
constraints must be equal to zero \cite{dirac}. The results of the
consistency requirement on the linearized constraints are given by
\begin{eqnarray}
\dot{\widetilde{H}} (t,{\bf k})&=&\int \frac{d^3k'}{(2\pi)^{3/2}}\left\{ \widetilde{H} (t,{\bf k}) , \widetilde{\mathcal{H}}(t,{\bf k'}) \right\} \nonumber \\
&=&k^2\left( \widetilde{{\sf p}} (t,{\bf k})+ 2k^2 \widetilde{\mathfrak{p}}^{\:l} (t,{\bf k})\right)=0 \label{kH2} ,
\end{eqnarray}
for the evolution of (\ref{kH}), and
\begin{eqnarray}
\dot{\widetilde{H}}_a (t,{\bf k})&=&\int \frac{d^3k'}{(2\pi)^{3/2}}\left\{ \widetilde{H}_a (t,{\bf k}) , \widetilde{\mathcal{H}}(t,{\bf k'}) \right\} \nonumber\\
&=& \frac{k^4 \widetilde{\mathfrak{h}}^{\:t}_a(t,{\bf k})}{4}=0
\label{kHi2}
\end{eqnarray}
for the time derivative of the linearized diffeomorphism constraint
(\ref{kHi}).  Equations (\ref{kH2}) and (\ref{kHi2}) define
secondary constraints.  These new constraints must be retroactively
placed into the action and, again the consistency requirement must
be checked.   The total Hamiltonian (constraint terms plus
$\widetilde{\mathcal{H}}$) is then transformed to
\begin{equation}
\widetilde{\mathcal{H}}_{tot}(t,{\bf k})  = \widetilde{\mathcal{H}}
-N\widetilde{H} -N^a \widetilde{H}_a  - M\dot{\widetilde{H}}
-M^a\dot{\widetilde{H}}_a  , \label{Htot}
\end{equation}
where the $(1)$ superscripts have now been completely suppressed and
$M(t,{\bf k})$, $ M^a(t,{\bf k})$ are the new Lagrange multipliers
to go with the secondary constraints $\dot{\widetilde{H}}(t,{\bf
k})$, $ \dot{\widetilde{H}}_a(t,{\bf k})$.  Normally in the
literature on linearized gravity $\cite{ADM}$, the secondary
constraints are automatically satisfied by the gauge conditions (or
time-slicing) chosen in the classical analysis. Only when one does
not make a coordinate or gauge choice, does one have these extra
constraints.

The story, however, does not end with (\ref{kH2}) and (\ref{kHi2}).
The time derivatives of these equations must again be calculated
and set equal to zero. The constraint-constraint terms in the
Poisson brackets must now be considered as they now longer commute.
Demanding that the double time
derivative of (\ref{kH}) be zero,
\begin{eqnarray}
\ddot{\widetilde{H}}(t,{\bf k}) =\int \frac{d^3k'}{(2\pi)^{3/2}} \left\{ \dot{\widetilde{H}}(t,{\bf
k}),\mathcal{H}(t,{\bf k}')\right\} +  \int \frac{d^3k'}{(2\pi)^{3/2}} N\left\{
\dot{\widetilde{H}}(t,{\bf k}),\widetilde{H}(t,{\bf k}')\right\} =0\nonumber \\
\end{eqnarray}
implies that $N$ is determined by this equation.  The $N^a$ are determined by evaluating the consistency condition
\begin{eqnarray}
\ddot{\widetilde{H}}_a(t,{\bf k})=\int \frac{d^3k'}{(2\pi)^{3/2}}\left\{ \dot{\widetilde{H}}_a(t,{\bf
k}) , \widetilde{\mathcal{H}}(t,{\bf k}') \right\} +\int \frac{d^3k'}{(2\pi)^{3/2}} N^b\left\{
\dot{\widetilde{H}}_a(t,{\bf k}) , \widetilde{H}_b (t,{\bf k}')
\right\}=0. \nonumber \\
\end{eqnarray}
Solutions for $M,M^a$ are in turn be found by recursively plugging in the
equations for $N,N^a$ into (\ref{kH}) and (\ref{kHi}).  The solutions
\begin{align}
M&=0 ,  &N=0, &&&\\
M^t_a&=-32 k^2  \widetilde{\mathfrak{h}}^{\:t}_a(t,{\bf k}),
&N_a=0&&&, \label{MNsols}
\end{align}
are uncovered by this final step. It should be noticed that the
Lagrange multipliers, except for $M^l$, have been determined by the
equations of motion.

The fact that the Lagrange multipliers for ($\widetilde{H}(t,{\bf
k})$, $\dot{\widetilde{H}}(t, {\bf k})$;
$\widetilde{H}^{\;t}_a(t,{\bf k})$, $ \dot{\widetilde{H}}\;^{t}_a(t,
{\bf k})$) are determined by the equations of motion leads us to
conclude that all constraints except $\widetilde{H}^{\; l}_a(t,{\bf
k})$ are \emph{second-class} constraints \cite{JRKQCS}. We
re-emphasize the point that this additional, second-class constraint
structure for linearized gravity only comes from the fact that we
are \emph{not} choosing a gauge.
 Conversely, $\widetilde{H}^l(t,{\bf k})$ is a first-class
constraint since its Poisson algebra is Abelian and its Lagrange
multiplier $M^l$ is not determined by the dynamics.

Before  promoting the canonical, phase-space variables $\{ {\bf
p}^{ab} , {\bf h}_{ab}; {\sf p}, {\sf h}; \mathfrak{p}^a ,
\mathfrak{h}_a \}$ to operators, one further step is taken to
rewrite the classical action.  An additional measure of convenience
may be afforded by rewriting the constraints so that they are each
functions of only one phase space variable. For a constrained
action, one is free to add and subtract linear combinations of the
constraints. Instead of using $\dot{\widetilde{H}}(t,{\bf
k}),\dot{\widetilde{H}}_a(t,{\bf k})$, the new constraints, the ones
we will actually quantize, are gained by adding and subtracting
linear combinations of $\{
\widetilde{H},\dot{\widetilde{H}},\widetilde{H}_a,
\dot{\widetilde{H}}_a \}$:
\begin{eqnarray}
\widetilde{\psi}(t,{\bf k})&\equiv& k\widetilde{{\sf h}}(t,{\bf k}) \label{psi}\\
\widetilde{\psi}_a(t, {\bf k}) &\equiv&  k\widetilde{\mathfrak{p}}^{\: t}_a(t,{\bf k})\\
\widetilde{\phi}(t,{\bf k})  &\equiv& \widetilde{{\sf p}}(t,{\bf k}) \\
\widetilde{\phi}_a(t,{\bf k})&\equiv& k^2 \widetilde{\mathfrak{h}}^{\: t}_a(t,{\bf k})\\
\widetilde{\varsigma}(t,{\bf k}) &\equiv& k^2
\widetilde{\mathfrak{p}}^{\: l}(t,{\bf k})\label{phi_a},
\end{eqnarray}
where the lone first class constraint is
$\widetilde{\varsigma}(t,{\bf k})$.  The fact that the phase-space
variables $\{ {\sf p}, {\sf h}; \mathfrak{p}^a , \mathfrak{h}_a \}$
are constrained is manifest in the final version of the total
Hamiltonian
\begin{equation}
\widetilde{\mathcal{H}}_{tot}(t,{\bf k}) =
\widetilde{\mathcal{H}}-N\widetilde{\phi} -N^a \widetilde{\phi}_a -
M\widetilde{\psi} -M^a\widetilde{\psi}_a -L\widetilde{\varsigma},
\label{Hgood}
\end{equation}
where each factor has had its functional dependence on $(t,{\bf k})$ suppressed.

To summarize, the above procedure found a classical action
describing linearized gravity most amenable to the Projection
Operator Method of quantization.  Insistence on a
gauge-independent classical theory and the perturbation scheme of (\ref{Gpert})-(\ref{Npert}) led to the necessity of
incorporating new constraints into the theory. These new constraints
came about as a result of the non-commutivity of the primary
constraints with the unconstrained Hamiltonian, leading to a second-class
designation for all but one of the original, primary
constraints---the longitudinal part of $\widetilde{H}_a$ that we
call $\widetilde{\varsigma}(t,{\bf k})$. While further investigation into this
algebraic structure is
 left to
future work, after uncovering all secondary constraints and adding
and subtracting linear combinations of these constraints with the
primary ones, a convenient, quantum-ready form of the classical
theory was found (\ref{psi})-(\ref{Hgood}).

\section{Quantization}

As a consequence of not choosing a gauge, any reduction of the
unconstrained theory to a physical set of quantities must be done
after quantization. The reduction is accomplished by use of a
projection operator ($\mathbb{E}$), which maps states in the
unconstrained Hilbert space ($ \vert \psi \rangle \in \mathfrak{H}$)
to states in the physical Hilbert space ($\vert \psi_P \rangle\in
\mathfrak{H}_P $), according to
\begin{equation} \mathbb{E} \vert \psi \rangle = \vert \psi_P \rangle .
\end{equation}
As a preliminary step, the Projection Operator is a function of the
sum of squares of the constraint operators ($\Sigma \Phi^2$) and
keeps their spectrum small, as indicated by
\begin{equation}
\mathbb{E}=\mathbb{E}[\Sigma \Phi^2 \leq \delta^2 ],
\end{equation}
for some small parameter $\delta >0$, whose exact value and behavior
depends on the constraints under consideration.  At this level,
$\Sigma \Phi^2$ could be a mixture of both first and second class
constraints.

 Quantum dynamics will then take place in this physical Hilbert
space using only the $\vert \psi_P \rangle$'s. In effect this
reduces $\mathfrak{H}$ to $\mathfrak{H}_P$, symbolically written as
\begin{equation}
\mathbb{E}\mathfrak{H}=\mathfrak{H}_P.
\end{equation}
Of course, this projection operator obeys all of the usual
properties of a projection operator, namely \begin{eqnarray}
\mathbb{E}^\dagger=\mathbb{E} \quad, \quad
\mathbb{E}^2=\mathbb{E}.\end{eqnarray} The coherent state matrix
elements of $\mathbb{E}$ define a reproducing kernel and are the key
to constructing the physical Hilbert space $\cite{JRKQCS,JRKearly}$.
This is done in a two part step---the reproducing kernel, like the
Projection Operator, is first regularized by the small parameter
$\delta$, and then $\delta$ is reduced to its smallest size
consistent with the spectrum of the constraint operators.

Generally, the reproducing kernel $\cite{JRKQCS, JRKearly}$
involving the Projection Operator is given by
\begin{eqnarray}
\langle \! \langle \: p',q'| \: p,q \: \rangle \!\rangle  \equiv
\frac{
 \langle \:p',q' | \mathbb{E} |\: p,q
\rangle }{ \langle 0,0|\mathbb{E} | 0, 0 \rangle} \label{regRK},
\end{eqnarray}
where $|p,q \rangle$ are suitable canonical coherent states.  The
re-scaling introduced by the reduction procedure, as indicated by
the denominator of (\ref{regRK}), simply involves the coherent state
matrix element of $\mathbb{E}$ for which all labels are zero.
Vectors in the physical Hilbert space are given by linear
superpositions of the reproducing kernel$\footnote{Vectors in the
original Hilbert space
 $\mathfrak{H}$ are given by similar expressions; except, in that case,
the reproducing kernel is the coherent state overlap function.}$
(\ref{regRK}), such as
\begin{equation}
\psi_P(p,q)= \sum^M_{m=1} \; \alpha_m \; \langle \! \langle \: p,q|
\: p_m,q_m \: \rangle \!\rangle. \label{RKvec2}
\end{equation}
For finite $M$, vectors of this sort form a dense set
$D_{\mathfrak{H}_P} \subset \mathfrak{H}_P$. In addition, inner
products of vectors in the dense set are determined by
\begin{equation}
(\psi, \phi)_P =  \sum_{m,n=1}^{M,N} \; \alpha^\ast_m \beta_n \;
\langle \! \langle \: p_m,q_m| \: p_n,q_n \: \rangle
\!\rangle.\label{RKip}
\end{equation}
The physical Hilbert space is completed by including the limit
points of all Cauchy sequences in the norm $\Vert \psi\Vert_P=(\psi,
\psi)^{1/2}_P$, completing the construction of the reproducing
kernel Hilbert space.

While the fundamental structure of the Projection Operator Method
treats first and second class constraints equally, the subspaces
over which the constraint operators act reduce in an entirely
different manner as $\delta \to 0$. In short, the reduction depends
on if the constraint operators in the argument of $\mathbb{E}$ are
first or second class. For the first class constraint of linearized
gravity, $\hat \varsigma$ has zero in its continuum, and a
Projection Operator involving only these constraints permits the
limit $\delta\to 0$. However, a $\delta$-dependent rescaling of the
reproducing kernel is incorporated into the projection operator in
order to prevent $\mathbb{E}\to 0$ as $\delta\to 0$. The rest of the
constraint operators, being second class in nature, do not have zero
in their continuum and a projection operator containing only these
constraints would forbid the limit of $\delta\to 0$. One might
expect the Projection Operator of linearized gravity to exhibit both
types of reduction since it contains both types of constraints.

Concerning the above statements: if we can find an appropriate
functional expression for the reproducing kernel, we can
characterize the Hilbert space in question.  Therefore, the
subsections to follow focus on finding suitable definitions of the
Projection Operator and coherent states so that we can determine the
exact form the reduced reproducing kernel assumes in linearized
quantum gravity.

\subsection{Operators and coherent states}

Quantization begins by the promotion of the $\Gamma$ coordinates
($p^{cd}(x), h_{ab}(y)$)  to sets of local, self-adjoint operators
$(\hat{p}^{cd}(x), \hat{h}_{ab}(y) )$, whose canonical commutation
relations$\footnote{Henceforth, operators will be represented by
hatted symbols such as $\hat{h}_{ab}$.  In addition, capital
superscripts on operators and labels referring to subspaces will be
suppressed in favor of their respective font denominations with the
exception being the lowercase $^t$ and $^l$ referring to the
transverse and longitudinal parts of the longitudinal vector
subcomponents.}$, in units where $\hbar=1$, read
\begin{equation}
\left[ \hat{h}_{ab}(x), \hat{p}^{cd}(x') \right]= i \delta^{c}_{\;\:(a}
\delta^{\;\:d}_{b)} \delta^3(x-x')\label{CCR},
\end{equation}
where parentheses denote index symmetrization.
As they stand in the above expression, the quantities
$\hat{h}_{ab}(x)$ and $\hat{p}^{cd}(y)$ are ill-defined as operators.

To introduce a regularization for the canonical operators,
let the vector ${\bf n}=\{ n_1, n_2, n_3\}$ label nodes on a finite, three-dimensional lattice of
volume $L^3$ so that band-limited Fourier coefficients for $\hat{f}_{ab}(x)$ may be found by
\begin{eqnarray}
 \hat{f}^{\bf n}_{ab}&=& \mathcal{V}^{-1/2}\int_{L^3} d^3x \; \me^{-i{\bf k^n}
\cdot {\bf x} }\hat{f}_{ab}(x), \label{latt}
\end{eqnarray}
 where $\mathcal{V}=(L/2\pi)^3$ and ${\bf k^n}\equiv 2\pi {\bf
n}/L$.
The operators themselves may then be regularized and approximated
by expansions based on these coefficients:
\begin{eqnarray}
\hat{f}_{ab}(x)&\equiv&\mathcal{V}^{-1/2}\sum_{{\bf n}\in \mathcal{I}}
 \; \hat{f}^{\bf n}_{ab}\me^{i {\bf k^{\bf n}}\cdot
{\bf x}}\label{klatt},
\end{eqnarray}
where the truncation of the sum is implicit on the left hand side.
 The Fourier series has been truncated in the above equation to include
only the lattice points in the set
\begin{equation}
\mathcal{I}= \left\{ {\bf n} | \; {\bf n} \in \mathbb{Z}^3 , \;
-N\leq n_a\leq N, \; {\bf n} \neq 0, N<\infty \right\},\label{ndefn}
\end{equation}
which means that the set of wave vectors ${\bf k^n}$ is band-limited
to
\begin{equation}
\mathcal{K}=\left\{ {\bf k^n}| \; {\bf k^n} \in \mathbb{R}^3 , \; \;
 0<|{\bf k^n}|\leq k_{max} , \; k_{max}= 2 \pi \sqrt{3} N/L \right\}. \label{kdefn}
\end{equation}

The self-adjoint requirement for $\hat{f}_{ab}(x)$ is met as long as
the reality condition $\hat{f}^{\dagger \; {\bf n}}_{ab}\equiv
\hat{\bar{f}}^{\bf n}_{ab}=\hat{f}^{-{\bf n}}_{ab}$ is implemented.
Lattice indices will reside in the middle of the alphabet, will be
kept raised and in bold font, and will not be automatically summed
if repeated. ({\bf Remark}: In nonperturbative quantum gravity, such
a simple Fourier series truncation is not available.  Instead of
summing over $\me^{i {\bf k^n}\cdot {\bf x}}$, the metric variables
may be smeared over real, orthonormal test functions of rapid
decrease \cite{JRKNC2}).

 The next step is to express each operator equation in terms of its Fourier
components.  Consequently, the momentum space CCR's$\footnote{Henceforth, operators will
be represented by hatted symbols such as $\hat{h}_{ab}$.  In
addition, superscripts on operators and labels will be suppressed
except for those involving longitudinal and transverse vector parts
of the longitudinal tensor subcomponents.}$ may be written as
\begin{equation}
\left[ \hat{\bar{h}}^{\bf m}_{ab}, \hat{p}_{cd}^{\bf n} \right]=
i\delta_{c(a} \delta_{b)d} \delta^{{\bf m,n}}\:\mathcal{V}. \label{alg}
\end{equation}
Using the above expression and
the expansion of (\ref{kdecomp}) with the $\widetilde{\:\:}$'s
replaced by $\hat{\:}$'s, the commutation relations of Appendix B
may be derived.  These commutation relations define the algebras of
each component in the factorized Hilbert space
$\mathfrak{H}=\mathfrak{H}_{TT} \otimes \mathfrak{H}_T \otimes
\mathfrak{H}_L$.

With the canonical commutation relations in hand, the ${\bf
k}$-space forms being found in Appendix B, the algebraic structure
of the quantum constraints may now be examined.  Promoting the
constraint functions in (\ref{psi})-(\ref{phi_a}) to operators, the
non-vanishing sector in the commutator algebra may still be found:
\begin{eqnarray}
\left[ \hat \phi (x) , \hat \psi (y) \right] &=& i \delta^3(x-y)\\
\left[ \hat{\phi}_a (x) , \hat{\psi}_b (y) \right] &=& i \delta^3(x-y)\delta_{ab} \label{QCCR}.
\end{eqnarray}
 As in the classical case, the constraint quantum constraint algebra can be classified as second class.

The CCR's given in (\ref{QCCR}) also allow canonical coherent states
for each subspace to be defined. Coherent states for this problem
admit a
 factorization in the form of
\begin{eqnarray}
\vert p, h \rangle &=&  \exp \left\{ -i (2\pi/L)^3\sum_{{\bf
n}\in \mathcal{I}}\; \left[\bar{h}^{\bf n}_{ab}
  \hat{ p}^{\bf n}_{ab}-\bar{p}^{\bf n}_{ab} \hat{ h}^{\bf n}_{ab }
   \right] \right\} \vert {\bf \eta} \rangle \nonumber \\ &\equiv&
 \vert   {\bf p , h }\rangle^{TT} \; \vert {\sf p , h }\rangle^T  \; \vert
 \mathfrak{p, h} \rangle^L, \label{cs1}
\end{eqnarray}
where a general fiducial vector $|{\bf \eta}\rangle$ has been used.
As the above expression for the coherent states in $\mathfrak{H}$ is
vital to our analysis, the rest of this section seeks to explain
each of the factors in ($\ref{cs1}$).

The first term in ($\ref{kdecomp}$) represents the
transverse-traceless components $(\hat{{\bf p}}_{ab},\hat{{\bf h}}_{ab})$, which
can in turn be expanded in terms of the two independent polarization states as
\begin{equation}
\hat{{\bf f}}^{\bf n}_{ab}=m_a m_b\hat{{\bf f}}^{\bf n}_{+} +
\bar{m}_a\bar{m}_b\hat{{\bf f}}^{\bf n}_- \label{TTexp}.
\end{equation}

This subspace has no associated constraint and represents the two
independent degrees of freedom in linearized gravity.  In terms of
only the transverse-traceless variables, the Hamiltonian density in
($\ref{kamel}$) may be interpreted as the operator
\begin{equation}
\hat{\mathcal{H}}_{TT}[{\bf p}, {\bf h}]= \hat{\bar{{\bf
p}}}_{ab}^{\bf n}\hat{{\bf p}}_{ab}^{\bf n}+ \frac{k^2}{4}
\hat{\bar{{\bf h}}}_{ab,c} ^{\bf n}\hat{{\bf h}}_{ab,c}^{\bf
n}.\label{TTkamel}
\end{equation}
The transverse-traceless coherent states $\cite{JRKFQO}$
then assume the form
%
%
\begin{equation}
\vert {\bf p, h} \rangle^{TT} = \exp \left\{ -i (2\pi/L)^3\sum_{{\bf
n}\in \mathcal{I}}\; \left[{\bf\bar{h}}^{\bf n}_{ab}
  \hat{{\bf p}}^{\bf n}_{ab }-{\bf \bar{p}}^{\bf n}_{ab} \hat{{\bf h}}^{\bf n}_{ab }
   \right] \right\} \vert O^{TT}
  \rangle, \label{TTCS}
\end{equation}
where $|O^{TT}\rangle$ is the ground state for (\ref{TTkamel}). The
coherent states in (\ref{TTCS}) comprise an overcomplete basis for
the transverse-traceless Hilbert space $\mathfrak{H}_{TT}$.  The
Weyl operator in ($\ref{TTCS}$) contains un-hatted quantities, which
serve dual purpose as being both smooth test functions on the
 lattice and the coherent state labels.

The transverse operators ($\hat{{\sf p}},\hat{{\sf h}}$) are represented
 by the second term in ($\ref{cs1}$) and act on states in the transverse
Hilbert space $\mathfrak{H}_T$.  To keep things sufficiently
general, coherent states, al\'{a} $\cite{JRKFQO}$ or
$\cite{beyond}$, in $\mathfrak{H}_T$ can be formed by the Weyl
operator acting on the transverse fiducial vector $|\eta^T\rangle$,
i.e.,
\begin{equation}
 \vert {\sf p, h }\rangle^T = \exp\left\{ -i(2\pi/L)^3\sum_{ {\bf n} \in \mathcal{I}}\;
 \left[\;{\sf \bar{h}}^{\bf n}\:
  \hat{{\sf p}}^{\bf n}  - {\sf \bar{p}}^{\bf n} \:\hat{{\sf h}}^{\bf n}\right]\right\}\vert
  \eta^T
  \rangle.  \label{TCS}
\end{equation}
However, the quantum structure of $\mathfrak{H}_T$ is restricted by
the constraint
\begin{eqnarray}
\hat{H}(x)=\sum_{ {\bf n} \in \mathcal{I}} k_{\bf n}^2\;\me^{i {\bf k^n}\cdot {\bf x}} \; \hat{{\sf h}}^{\bf n} \label{QH_0},
\end{eqnarray}
where $k_{\bf n}^2\equiv  |{\bf k^n}|^2= k^{\bf n}_a k^{\bf n}_a $.

The other constrained Hilbert space is the longitudinal (vector)
Hilbert space $\mathfrak{H}_L$. As with the case for $|\eta^T\rangle$,
 the fiducial vector of the longitudinal coherent states,
$|\eta^L\rangle$, is the longitudinal fiducial vector.  These fiducial
vectors, left as general states for now, will be determined in the next
subsection.  The Weyl operator version of the coherent states in $\mathfrak{H}_L$
is conveniently derived when
\begin{equation}
\mathfrak{f}^t_a\equiv\frac{ \bar{m}_a\mathfrak{f}^t_1 + m_a \mathfrak{f}^t_2}{2}
\end{equation}
is inserted into the definition of (\ref{cs1}).  The operator
version of the expansion of (\ref{kdecomp}), applied to (\ref{cs1}),
then allows us to extract
\begin{align}
\vert \mathfrak{p}, \mathfrak{h} \rangle^L &= \exp \left\{-i
(2\pi/L)^3\sum_{ {\bf n} \in \mathcal{I}}\;4k^2_{\bf n}\left[
   \mathfrak{\bar{h}}^{t\; {\bf n}}_a \mathfrak{\hat{p}}^{t\; {\bf n}}_a-
\mathfrak{\bar{p}}^{t\;{\bf n}}_a \mathfrak{\hat{h}}^{t\;{\bf n}}_a
+k^2_{\bf n} \left(\mathfrak{\bar{h}}^{l\; {\bf n}}
\mathfrak{\hat{p}}^{l\; {\bf n}}-
  \mathfrak{\bar{p}}^{l\;{\bf n}}\mathfrak{\hat{h}}^{l\;{\bf n}}\right)\right]\right\} \vert \eta^L \rangle \nonumber \\
&= \exp \left\{-i
(2\pi/L)^3\sum_{ {\bf n} \in \mathcal{I}}\;4k^2_{\bf n}\left[
   \mathfrak{\bar{h}}^{t\; {\bf n}}_a \mathfrak{\hat{p}}^{t\; {\bf n}}_a-
\mathfrak{\bar{p}}^{t\;{\bf n}}_a \mathfrak{\hat{h}}^{t\;{\bf n}}_a\right] \right\} |\eta^t\rangle \nonumber \\
&\; \quad\otimes  \exp \left\{-i
(2\pi/L)^3\sum_{ {\bf n} \in \mathcal{I}}\;4k^4_{\bf n}\left[
   \mathfrak{\bar{h}}^{t\; {\bf n}}_a \mathfrak{\hat{p}}^{l\; {\bf n}}_a-
\mathfrak{\bar{p}}^{l\;{\bf n}}_a \mathfrak{\hat{h}}^{l\;{\bf
n}}_a\right] \right\} |\eta^l\rangle \nonumber\\
&= |\mathfrak{p}, \mathfrak{h}\rangle^t \;|\mathfrak{p},
\mathfrak{h}\rangle^l\label{LCS}
\end{align}
such that $|\eta^t\rangle\otimes|\eta^l\rangle=|\eta^L\rangle$.  This means our starting fiducial vector $|{\bf \eta}\rangle$ may be expressed in factorized, direct product form as
\begin{equation}
|{\bf \eta}\rangle=|O^{TT}\rangle \otimes |\eta^T\rangle \otimes (|\eta^t\rangle \otimes
|\eta^l\rangle ) \label{fiducial}.
\end{equation}

To give a brief overview of the above discussion, convenient forms
for the coherent states and operators were found. In promoting the
position space versions of  (\ref{psi})-(\ref{phi_a}) to
self-adjoint constraint operators, a second-class constraint
structure is unavoidable. Kinematical coherent states in
$\mathfrak{H}$ are given by ($\ref{cs1}$), defined as a direct
product of states in $\mathfrak{H}_{TT}$, $\mathfrak{H}_T$, and
$\mathfrak{H}_L$.  Each of the (canonical) coherent states
provides an overcomplete basis for their corresponding subspace in
$\mathfrak{H}$. This unconstrained Hilbert space may be expressed in
direct product form as
\begin{equation}
\mathfrak{H}= \mathfrak{H}_{TT}\otimes \mathfrak{H}_T \otimes
\mathfrak{H}_L.
\end{equation}
Of the three subspaces, the transverse and longitudinal Hilbert
spaces are constrained when acted upon by the Projection Operator.  We seek to examine
how this works in the next section.
%
\subsection{The reduced reproducing kernels}

With the quantum operators, constraints, and coherent states
defined, the only item left remaining to do is to determine the
reduced reproducing kernel, as this will uncover the nature of the
physical Hilbert space $\mathfrak{H}_P$. The projection operator is
constructed so that $\mathbb{E}=\mathbb{E}\left[ \Sigma \Phi^2 \leq
\delta^2\right] $, where $\delta$ is a small, but nonzero,
parameter. As introduced in earlier works,  $\Sigma \Phi^2$ can be
defined as
\begin{eqnarray}
\Sigma \Phi^2 &=&\sum_{ {\bf n} \in \mathcal{I}} \left(| \hat{\psi}^{\bf n}|^2
+ |\hat{\phi}^{\bf n}|^2\right)+\sum_{ {\bf n} \in \mathcal{I}}\left(|\hat{\psi}_a^{\bf n}|^2
 + |\hat{\phi}^{\bf n}_a|^2 \right) +\sum_{ {\bf n} \in \mathcal{I}}| \hat{\varsigma}^{\bf n}|^2
\nonumber \\
&=&\sum_{ {\bf n} \in \mathcal{I}}
\left( |\hat{{\sf p}}^{\bf n}|^2 +   k^2_{\bf n}|\hat{{\sf h}}^{\bf n}|^2 \right) \nonumber
\\ &\;& \; + \sum_{ {\bf n} \in \mathcal{I}}k^2_{\bf n} \left( |
\hat{\mathfrak{p}}_a^{t\; {\bf n}}|^2 + k^2_{\bf n}
|\hat{\mathfrak{h}}_a^{t\; {\bf n}}|^2 \right) +\sum_{ {\bf n} \in
\mathcal{I}}k^4_{\bf n}|\hat{\mathfrak{p}}^{l\; {\bf n}}|^2 ,
\label{SigmaPhifull}
\end{eqnarray}
where position and momentum operators conjugate to one another have been suggestively
 grouped together with parentheses.  The constraint operators in (\ref{SigmaPhifull}) are quantized from the constraint functions in (\ref{psi})-(\ref{phi_a}).
The first two parenthetical terms in (\ref{SigmaPhifull}) can be interpreted heuristically as, being quadratic in
$\{ \hat{{\sf p}}^{\bf n},\hat{{\sf h}}^{\bf n} \}$ and
$\{ \hat{\mathfrak{p}}^{t\;\bf n}, \hat{\mathfrak{h}}^{t\;\bf n}\}$, a sum of harmonic-oscillator
Hamiltonians
on the momentum lattice in both the transverse and longitudinal subspaces.  With this insight,
it is anticipated that $\delta$ cannot be taken to zero in any limit without $\mathbb{E}$
 vanishing everywhere.
Thus, $\delta$ must have a positive-valued minimum so as to capture
only the ground state eigenvalue of each oscillator, a situation
typical in second-class systems with the Projection Operator Method
\cite{JRKQCS}. However preventing $\delta \to 0$ for the spectrum of
the $| \hat{\varsigma}^{\bf n}|^2$ operator amounts to a violation
of the quantum constraint.
 Therefore, the appropriate $\delta$ to use for this mixed case of first and second class constraints is
\begin{equation}
\delta^2\to \delta^2 +3\left( \frac{2\pi}{L}\right)^3\sum_{{\bf n}
\in \mathcal{I}}\hbar k_{\bf n}.
\end{equation}

 The Projection Operator is formed by taking into account the fact that $\mathfrak{H}$ is formed by a direct product of Hilbert spaces
\begin{eqnarray}
\mathfrak{H}&=&\mathfrak{H}_{TT}\otimes \mathfrak{H}_T \otimes \mathfrak{H}_L \\
&=& \mathfrak{H}_{TT}\otimes \mathfrak{H}_T \otimes \left(\mathfrak{H}_t\otimes \mathfrak{H}_l
\right),
\end{eqnarray}
where the second line above exhibits the independence of the vector degrees of freedom of $\mathfrak{H}_L$. This suggests that we can split the projection operator according to
\begin{align}
\mathbb{E}&=\mathbb{E} \left[ \Sigma \Phi^2 \leq \delta^2 \right] \label{E}\\
 &=\mathbb{E} \left[ \Sigma (|\hat{{\sf p}}|^2 + k^2|\hat{{\sf h}}|^2) +
\Sigma ( |\hat{\mathfrak{p}}|^2+ k^2|\hat{\mathfrak{h}}|^2 )\leq
\delta^2+3 \Sigma\hbar k\right] \label{E2}
 \\ &= \mathbb{E}^T\left[\Sigma (|\hat{{\sf p}}|^2 + k^2|\hat{{\sf h}}|^2) \leq \Sigma \hbar k\right] \otimes  \mathbb{E}^t \left[ \Sigma k^2(| \hat{\mathfrak{p}}^t|^2+ k^2|\hat{\mathfrak{h}^t}|^2 )
\leq 2\Sigma \hbar k
 \right] \nonumber\\
&\; \quad \quad  \otimes \mathbb{E}^l \left[\Sigma k^4
|\hat{\mathfrak{p}^l}|^2 \leq \delta^2  \right] , \label{E3}
\end{align}
using the shorthand  $\Sigma$ for $\mathcal{V}^{-1}\sum_{{\bf n}\in \mathcal{I}}$ . Since
the operator arguments of $\mathbb{E}^T$ and $\mathbb{E}^t$ may be
interpreted as the Hamiltonians of harmonic oscillators, the
transition from (\ref{E2}) to (\ref{E3}) is made by assigning the
appropriate multiplicative factors of $\Sigma \hbar k$ to capture
the associated ground state of each oscillator.

More specifically, the transverse Projection Operator $\mathbb{E}^T$
in (\ref{E3}) confines the eigenvalues of $\Sigma (\hat{{\sf p}}^2 +
k^2\hat{{\sf h}}^2) $ to being in their ground state values for each
point on the lattice in $\mathfrak{H}_T$. Its counterpart,
$\mathbb{E}^t$ , similarly restricts the two degrees of freedom per
lattice point in $\mathfrak{H}_t$. The only nonzero states which
will meet the inequality conditions of $\mathbb{E}^T$ and
$\mathbb{E}^t$ in (\ref{E3}) are the ground states of each operator.
We use the familiar projection operator notation of quantum
mechanics to abbreviate (\ref{E3}) by
\begin{eqnarray}
\mathbb{E}= | O^T\rangle\langle O^T |\otimes | O^t\rangle \langle
O^t| \otimes \mathbb{E}^l \left[\Sigma |\hat{\varsigma}|^2
\leq \delta^2 \right] \label{E4}.
\end{eqnarray}

The projection operator $\mathbb{E}^l$ is constructed in a
fundamentally different manner from those of the $\mathfrak{H}_T$
and $\mathfrak{H}_t$ subspaces, as the $\hat{\varsigma}^{\bf n}$
constraints have zeros in their continuous spectrum. The projection
operator $\mathbb{E}^l$ can be formally written as
\begin{eqnarray}
\mathbb{E}^l&=&\mathbb{E}^l \left[ \Sigma |\hat{\varsigma}|^2 \leq \delta^2 \right] \nonumber\\
 &=&\int^{\infty}_{-\infty} d\lambda \; \exp \left[ -i\lambda  \Sigma |\hat{\varsigma}|^2  \right]
 \; \frac{\sin(\delta^2 \lambda)}{\pi \lambda} .\label{projgen}
\end{eqnarray}
 A suitable limit where $\delta\to 0$ is reserved for a later
stage.

We limit the discussion on $\mathbb{E}^l$ here to formal arguments
for good reason.  It is well-known that $\mathbb{E}^l$ is uniquely
determined by its coherent state matrix elements \cite{JRKCS}. Since
the reproducing kernel is defined by the very same coherent state
matrix elements of $\mathbb{E}$ (\ref{regRK}) and provides Gaussian
smoothing owing to the test function properties of the coherent
states, we insist on the coherent state matrix elements of
$\mathbb{E}$ as being rigorously-defined mathematical quantities
\cite{JRKCS}.

Knowing that the Projection Operator factorizes according to
(\ref{E4}), with the definition of (\ref{regRK}) the reduced
reproducing kernel can be given the functional forms of the
transverse and longitudinal reduced reproducing kernels,
\begin{align}
\langle \! \langle \: p',h'&| \: p,h \: \rangle \!\rangle \equiv
\langle\!\langle {\bf p',h'} \vert {\bf p, h} \rangle \!\rangle^{TT}
\; \langle\!\langle {\sf p',h'} \vert {\sf p,h} \rangle\!\rangle^T
\; \langle\!\langle  \mathfrak{p}', \mathfrak{h}' \vert
\mathfrak{p},\mathfrak{h} \rangle\!\rangle^L \nonumber \\
&=  \langle{\bf p',h'} \vert {\bf p, h} \rangle^{TT}  \; \frac{
\langle \:{\sf p',h'} | \mathbb{E}^T |\: {\sf p,h} \rangle }{
\;\langle  0, 0 | \mathbb{E}^T | 0,0 \rangle}\;\frac{ \langle
\:\mathfrak{p}'^t ,\mathfrak{h}'^t |
 \mathbb{E}^t |\mathfrak{p}^t,\mathfrak{h}^t
\rangle }{ \;\langle {\bf 0},{\bf 0} | \mathbb{E}^t | {\bf 0}, {\bf
0} \rangle} \nonumber  \lim_{\delta \to 0}\;\frac{ \langle
\:\mathfrak{p}'^l,\mathfrak{h}'^l |
 \mathbb{E}^l |\mathfrak{p}^l,\mathfrak{h}^l
\rangle }{ \;\langle  0, 0 | \mathbb{E}^l |  0, 0 \rangle}\\ &=
\langle{\bf p',h'} \vert {\bf p, h} \rangle^{TT} \mathcal{K}_2 [{\sf
p',h'} ;{\sf p,h}] \; \mathcal{K}_2
[\mathfrak{p}'^t,\mathfrak{h}'^t;\mathfrak{p}^t,\mathfrak{h}^t]
\mathcal{K}_1[\mathfrak{p}'^l,\mathfrak{h}'^l;\mathfrak{p}^l,\mathfrak{h}^l]
\label{Kdelta}
\end{align}
The functional $ \mathcal{K}_2  $
represents the functional for reproducing kernels for second-class
constraints in (\ref{Kdelta}), while $ \mathcal{K}_1 $ represents the reproducing kernels for their
first-class counterparts.  Each of these reproducing kernels can be
shown to be the reproducing kernels of one-dimensional Hilbert
spaces.

This fact
 may not be immediately obvious.
Writing down any member of a dense set $\Psi\in D_{\mathfrak{H}_T}$
of the transverse Hilbert space $\mathfrak{H}_T$ leads to
\begin{eqnarray}
\Psi[{\sf p},{\sf h}]&=&\sum_m  \beta_m  \; \langle
{\sf p},{\sf  h} |{\sf p}_m, {\sf  h}_m
\rangle \nonumber \\
&=&\sum_m \beta_m \mathcal{K}_2 [ {\sf p},{\sf h};{\sf p}_m,{\sf h}_m] \nonumber \\
&=& \sum_m \beta_m \langle \:{\sf p},{\sf h} |O^T\rangle \langle
O^T|{\sf p}_m,{\sf h}_m\rangle
\\ &\to & \Psi_0 [{\sf p},
{\sf h}]\sum_m \beta_m \me^{-\sum \left[\frac{|{\sf p}_m|^2}{4k} +\frac{k|{\sf h}_m|^2}{4}\right]}
 \propto \Psi_0
[{\sf p}, {\sf h}], \label{Psi}
\end{eqnarray}
The $\Psi_0$ in the above equation is the ground state
representative in $\mathcal{H}_T$:
\begin{equation}
\Psi_0[{\sf p}, {\sf h}]= \langle \! \langle  {\sf p}
, {\sf h} | 0, 0 \rangle\!\rangle = \me^{-\sum \left[\frac{|{\sf p}|^2}{4k} +\frac{k|{\sf h}|^2}{4}\right]} .
\end{equation}
The fact that every vector in the dense set $D_{\mathfrak{H}_T}$ is
proportional to the ground state $\Psi_0$ leads to the conclusion
that the transverse reproducing kernel characterizes a
one-dimensional Hilbert space.  The same analysis also holds for
$\Psi \in D_{\mathfrak{H}_t}$ given by expansions of $\mathcal{K}_2
[\mathfrak{p}'^t,\mathfrak{h}'^t;.\mathfrak{p}^t,\mathfrak{h}^t]$. A
similar result follows for the reduction of $\mathcal{K}_1$, the
specifics of which are included in Appendix B.

Now that it has been demonstrated that the factors $\mathcal{K}_2$  and $\mathcal{K}_1$ in
(\ref{Kdelta}) are extraneous, one may
also choose to absorb these factors into a re-definition of the
coherent states \cite{JRKQCS}. With this operation, the reproducing kernel
takes on the simplified form of
\begin{equation}
  \langle \! \langle \: p',h'| \: p,h \: \rangle
\!\rangle=\langle{\bf p',h'} \vert {\bf p, h} \rangle^{TT}.
\label{end}
\end{equation}
In representing the reduced reproducing kernel in (\ref{end}), we
have suppressed the trivial nature of the
 labels $\{ \mathfrak{p}, \mathfrak{h}; {\sf p}, {\sf h}\}$.

The true dynamical degrees of freedom left reside entirely in the
transverse-traceless components of the metric and momentum fields,
as expected. The Hilbert space $\mathfrak{H}_{TT}$ is the physical
Hilbert space $\mathfrak{H}_P$, and general vectors in this space
are given by expansions of the (\ref{end}) reproducing kernel,
producing vectors of the form of (\ref{RKvec2}). The calculations
that went into finding ($\ref{end}$) show how this was achieved
without choosing any lapses, shifts, or time representations;
instead, a projection operator was used to enforce the quantum
constraints.

\section{Gravitonic states in the physical Hilbert space}

The reproducing kernel formalism not only provides a useful vehicle
to describe reduction of the original Hilbert space, but for the
problem under consideration, it also can be used to build a
functional Fock space representation of the physical Hilbert space.
To start with, the ground state functional representative in the
physical Hilbert space may be written in the continuum limit of
$L\to \infty$ and $N \to \infty$ as
\begin{eqnarray}
\Psi_0[{\bf p},{\bf h}]&\equiv&\langle {\bf p, h}  \vert 0 \rangle^{TT} \label{eq90} \\
&=&\exp \left[ - \frac{1}{2}\int d^3k \frac{1}{\omega({\bf k})}
\left(
  \vert {\bf p}_{ab }\vert^2 + \frac{\omega({\bf k})^2}{4} \vert {\bf h}_{ab }
  \vert^2\right) \right] , \nonumber
\end{eqnarray}
where $\omega= \vert {\bf k} \vert$.  Using ($\ref{TTexp}$), the complex
modulus-squared of each label may be expanded in terms of its components as
\begin{eqnarray}
|{\bf h}_{ab }|^2=| h_+ ({\bf k})|^2 + | h_-({\bf
k})|^2, \quad |{\bf p}_{ab }|^2=|p_+ ({\bf k})|^2 +
| p_-({\bf k})|^2,
\end{eqnarray}
where the $+$ and $-$ subscripts denote the two different graviton
polarizations.  This means that the ground state functional can also be
seen as an independent functional for each polarization, or
\begin{eqnarray}
\Psi_0[{\bf p},{\bf h}]&=& \Psi_{0+}[ p_+,  h_+] \:
\Psi_{0-}[ p_-,  h_-],
\end{eqnarray}
where the functional $\Psi_{0\pm}[ p_\pm,  h_\pm]$ is defined in
the same way as ($\ref{eq90}$), but with $({\bf p}_{ab},{\bf
h}_{ab})$ replaced by $ ( p_\pm, h_\pm)$.

To populate Fock space with gravitonic states, one needs to know
what the creation and annihilation operators are in the appropriate
coherent state representation \cite{JRKFQO}. The most convenient way
to do this is to introduce the complex label ${\bf z}({\bf k})$, the
components of which are given by
\begin{equation}
 z_\pm({\bf k})=\frac{ \sqrt{\omega}}{2}h_\pm ({\bf k}) + \frac{i}{\sqrt{\omega}} p_\pm ({\bf
 k}),
\end{equation}
for each polarization state. Using this new label, ($\ref{eq90}$)
can be expressed as
\begin{equation}
\Psi_0[{\bf z}]=\exp \left[ - \frac{1}{2}\int d^3k \left( | z_+({\bf
k})|^2+ | z_-({\bf k})|^2 \right)\right].
\end{equation}
The annihilation operators for both the $+$ and $-$ polarization
state appear then as
\begin{eqnarray}
a_\pm({\bf k})=\frac{z_\pm({\bf k})}{2}
 +\frac{\delta}{\delta  z^\ast_\pm({\bf k})} .
\end{eqnarray}
Likewise, the creation operator for each state is
\begin{eqnarray}
a^\dagger_\pm({\bf k})=\frac{z^\ast_\pm({\bf k})}{2}
 -\frac{\delta}{\delta  z_\pm({\bf k})}.
\end{eqnarray}
These operators act on the ground state $\Psi[{\bf z}]$ to give
\begin{eqnarray}
a_\pm({\bf k})\Psi_0[{\bf z}]&=&0, \\
a^\dagger_\pm({\bf k}_1)\dots a^\dagger_\pm({\bf k}_l)\Psi_0[{\bf
z}] &=& z^\ast_\pm({\bf k}_1)\dots  z^\ast_\pm({\bf
k}_l)\Psi_{0}[{\bf z}].
\end{eqnarray}

\section{Conclusion}

This work examined the application of the Projection Operator Method
to the theory of linearized gravity.  In the classical version of
the theory, perturbing around a flat background and insisting on
gauge independence led to a set of partially second-class, classical
constraints. In fact, only one degree of freedom per lattice point
was constrained in a first class manner, corresponding to a gauge
choice which could have been made.  While the emergence of the
second-class nature of the constraints at the classical level
certainly seems novel and merits further investigation, the main
thrust of this work involves the demonstration that the Projection
Operator can unambiguously quantize an infinite number of mixed
 first and second-class constraints.

The core of the quantum analysis was the calculation of the
reproducing kernels.  These results showed explicitly how the
Projection Operator, though it contained a factorizable product of
projection operators, reduced all constrained subspaces to
independent copies of the one-dimensional Hilbert space of complex
numbers ($\pmb{1}_\mathbb{C}$). Symbolically, leaving needed rescaling as implicit, this may be expressed
as
\begin{eqnarray}
\mathfrak{H}_P&= & \mathfrak{H}_{TT} \otimes \mathbb{E}^T
\mathfrak{H}_T \otimes (\mathbb{E}^t
 \mathfrak{H}_t \otimes \mathbb{E}^l \mathfrak{H}_l) \nonumber \\
&= &\mathfrak{H}_{TT} \otimes \pmb{1}_\mathbb{C} \otimes( \pmb{1}_\mathbb{C}\otimes \pmb{1}_\mathbb{C} ) \label{preiso}\\
&\leadsto& \mathfrak{H}_{TT}.  \label{postiso}
\end{eqnarray}
Here,
 we have introduced the symbol $\leadsto$ as meaning that
a bijection can be found relating the second and third lines,
($\ref{preiso}$) and ($\ref{postiso}$).  This result agrees with the
well-known result that the transverse-traceless degrees of freedom
are the only degrees of freedom involved in quantum dynamics. This
reduction mirrors the result obtained recently by Dittrich and
Thiemann using the Master Constraint Programme \cite{Thiemann}.

To further illustrate the connection to previous results, equation
($\ref{eq90}$) may be compared with prior work by Kucha\v{r}
$\cite{kook}$.  In his version, the ground state functional for
linearized gravity is given by
\begin{equation}
\Psi_0[h^{TT}]= \mathcal{N} \exp \left[- \frac{1}{4} \int d^3k \;\;
\omega({\bf k})\; \left\vert h^{TT}_{ab }\right\vert^2
\right], \label{kook}
\end{equation}
the two physical degrees of freedom in the theory residing within
the tensor $ h^{TT}_{ab}$.  For us these same degrees of freedom are embedded in
$\{{\bf p},{\bf h}\}$.  Here $\mathcal{N}$ is a formal
normalization constant.  It is clear that in passing to a
representation involving only the ${\bf h}_{ab}({\bf k})$,
($\ref{eq90}$) results in an expression identical to ($\ref{kook}$).

In essence, we can say that the Projection Operator Method,
combined with the reduced reproducing
 kernel calculations, has shown that the transverse and
longitudinal degrees of freedom completely decouple from linearized
gravity after quantization.  Kucha\v{r} discovered the same dynamics
as a result of  traditional canonical quantum gravity techniques: by
using the so-called extrinsic time representation, a gauge choice,
and embedding the quantum dynamics into the WdW equation. Our result
for the ground state functional is similar to what one would expect
from using the Wheeler-DeWitt equation, however we remark again that
no gauge has been chosen in our approach.

\appendix
\section{Appendix: Simple Toy Model} \label{A}

To help  clarify the way in which the Projection Operator
reduces the kinematical Hilbert space to the physical Hilbert space for a second-class system of constraints,
we have chosen to include an example problem analogous to the situation of
gauge-independent linearized gravity.  Consider a system in
a flat phase space, coordinatized by ($ p_a,
q^a$), $a=1,2,3$, such that
\begin{equation}
\{ q^b, p_a \} = \delta_a^b
\end{equation}
Let this system be described by the constrained action
\begin{eqnarray}
S[p,q]=\int dt \left[ p_a \dot{q}^a- H(p,q)
-\lambda^A p_A \right] \label{toyaction1},
\end{eqnarray}
where $A=2,3$.  The unconstrained Hamiltonian is then given by
\begin{equation}
H(p,q)=  \frac{p^2_1}{2} +  \frac{q^2_1}{2} + \sum^3_{J=2}\left(
\frac{p^2_J}{2} -\frac{q_J^2}{2}\right). \label{toyH}
\end{equation}

Evolution of the constraints is directly analogous to the reduction
involving $\mathfrak{H}_t$ in gauge-independent linearized gravity.
The constraint $p_A$ in (\ref{toyaction1}) most similar to the
constraint piece involving $\mathfrak{p}^t_a$ in (\ref{Hilin}),
while the Hamiltonian in (\ref{toyH}) is of the same basic form as
(\ref{Htot}). Calculating the evolution of $p_A$, one finds
\begin{equation}
\dot{p}_A=\{ p_A,H\} =q_A\equiv 0 \ \label{p_A=0},
\end{equation}
meaning that we must consider $q_A$ as a secondary constraint.

Placing primary and secondary constraints back in the action, we get
\begin{equation}
S[p,q]=\int dt \left[ p_a \dot{q}^a- H(p,q) -\lambda^A
p_A  -\mu^B q_B\right] \label{toyaction2},
\end{equation}
where the set of Lagrange multipliers $\{ \mu_A,\lambda_A \}$ are
determined by the equations of motion.  Now (\ref{toyaction2}) is
used to recursively calculate the evolution of the constraints.
After again insisting on the vanishing of the time-derivatives of
all constraints, the solution
\begin{eqnarray}
\lambda_A=q_A=0=  -p_B =\mu_B\label{lambdasol}
\end{eqnarray}
is obtained as an end result.

Foregoing further analysis and physical interpretation of the
classical solution, we proceed to the quantized version of the
theory.  Observing that the classical constraint algebra,
\begin{eqnarray}
\left\{ q^B, p_A \right\}=\delta_A^B ,
\end{eqnarray}
does not vanish on the constraint hypersurface,  we come to the
conclusion that this is a second-class system of constraints. Since
the constraints are just the canonical variables, promotion of the
phase-space coordinates to self-adjoint, irreducible operators via
\begin{equation}
\{ q^a, p_b \} = \delta^a_b \mapsto -i[Q^a,
P_b ]/\hbar,
\end{equation}
 implies the promotion of the constraints to self-adjoint operators.  The fact that
the constraint commutator algebra is by definition
\begin{equation}
\left[Q^B, P_A\right] = i\hbar,
\end{equation}
means that there is still a second class constraint structure for the quantum analysis.

Construction of the Projection Operator is straightforward. This operator is defined as
\begin{eqnarray}
\mathbb{E}\equiv \mathbb{E}[ \Sigma\Phi^2 \leq
\delta^2]=\mathbb{E}[\Sigma( P_A^2 + Q_A^2)\leq 2\hbar ].
\end{eqnarray}
Since $\Sigma \Phi^2$ is clearly the Hamiltonian operator for a
unit-frequency, two-dimensional harmonic oscillator, $\delta^2$ may
be chosen as $2\hbar$ to capture only the ground state and prevent
$\mathbb{E}$ from vanishing. Therefore, working in the energy
eigenbasis the Projection Operator may be simply written as
\begin{equation}
\mathbb{E}[\Sigma (P_A^2 + Q_A^2)\leq 2\hbar] = | O_2 ,O_3\rangle
\langle O_2,O_3|,\label{toyE}
\end{equation}
where $ | O_2 ,O_3\rangle \langle O_2,O_3|$ is the ground state for
the $J=2,3$ degrees of freedom.  Introducing  $ | {\bf O} \rangle$
as the ground state for all three degrees of freedom, the coherent
states over the unconstrained Hilbert space are defined as
\begin{eqnarray}
|{\bf p ,q} \rangle& =&  \me^{i (p_a Q^a-q^a P_a) } | {\bf O} \rangle\\
&\equiv& | p_1, q_1 \rangle \otimes |p_2 , q_2 \rangle \otimes |p_3,q_3\rangle
 \label{toycs},
\end{eqnarray}
such that $q_a$ and $p_a$ are the expectations of the $Q_a$ and $P_a$ operators respectively.
Using (\ref{toyE}) with (\ref{toycs}), the reproducing kernel is given by
\begin{eqnarray}
\langle\!\langle {\bf p',q'} | {\bf p,q}\rangle\!\rangle &=&\langle
{\bf p',q'}| O\rangle \langle O| {\bf p,q}\rangle
 \label{toy2}\\
&=& \mathcal{K}_0[p'_1,q'_1;p_1,q_1]\;
\mathcal{K}_2[p'_A,q'_A;p_A,q_A], \label{toy3}
\end{eqnarray}
where $\mathcal{K}_2[p'_A,q'_A;p_A,q_A]= \me^{-(p'^2_A+q'^2_A)/4
-(p^2_A+q^2_A)/4 }$.  This equation for $\mathcal{K}_2$ should be
compared with $\mathcal{K}_0[p'_1,q'_1;p_1,q_1]$, which is the
conventional coherent state overlap
\begin{eqnarray}
\mathcal{K}_0[p'_1,q'_1;p_1,q_1]&=&  \langle p'_1,q'_1  | p_1,q_1 \rangle \nonumber\\
&=& \me^{-(p_1-p'_1)^2/4-(q_1-q'_1)^2/4 +i(p_1q'_1-p'_1q_1)/2 }.
\end{eqnarray}

The physical Hilbert space characterized by the reproducing kernel
in (\ref{toy3}) may be expressed as
\begin{equation}
\mathfrak{H}_P=\mathfrak{H}_1\otimes\mathbb{E}\left(\mathfrak{H}_2\otimes
\mathfrak{H}_3\right). \end{equation}
 Vectors $\Psi \in D_{\mathfrak{H}_P}$ for a dense set $D_{\mathfrak{H}_P} \subset \mathfrak{H}_P$  can
be written as
\begin{align}
\Psi[{\bf p},{\bf q}]&= \sum_m \; \beta_m \langle\!\langle {\bf p},
{\bf q}|{\bf p}_m, {\bf q}_m \rangle\!\rangle \nonumber \\
&=\sum_m \; \beta_m
\mathcal{K}_0[p_1,q_1;p_{1m},q_{1m}]\mathcal{K}_2[p_A,q_A;p_{Am},q_{Am}]\nonumber \\
&= \sum_m \left(\beta_m
\me^{-(p^2_{Am}+q^2_{Am})/4}\me^{-(p^2_A+q^2_A)/4}\right)\;
\mathcal{K}_0[p_1,q_1;p_{1m},q_{1m}]\nonumber
\\&= \sum_m \beta'_m \;\mathcal{K}_0[p_1,q_1;p_{1m},q_{1m}]\label{toy4}
\end{align}
where $\mathcal{K}_2$ has been absorbed into the complex
coefficients $\beta'_m$. The reduction depicted in (\ref{toy4})
 shows how the structure of $\mathcal{H}_P$ is only dependent on the
single $(p_1,q_1)$ degree of freedom.

\section{Appendix: Theorem on the reduction of a first-class reproducing kernel using coherent states} \label{B}

To show the collapse of the constrained Hilbert space
$\mathfrak{H}_l$ into a one-dimensional Hilbert space, we shall
prove the following

{\bf Theorem}: The Hilbert space $\mathfrak{H}_l$ characterized by
$\mathcal{K}_1 [\mathfrak{p}'^l,\mathfrak{h}'^l;.\mathfrak{p}^l,\mathfrak{h}^l]$ is a one-dimensional Hilbert space.

{\bf Proof}:  The direct proof to follow involves the lattice setup
of (\ref{ndefn}) and (\ref{kdefn}) and the properties of coherent
states contained in , e.g., \cite{JRKCS} and \cite{JRKFQO}.  The
longitudinal vector projection operator, extracted from (\ref{E3}),
is given by
\begin{eqnarray}
\mathbb{E}^l=\mathbb{E}^l[ \Sigma |\hat{\varsigma}|^2\leq\delta^2].
\end{eqnarray}
It pays to examine the argument of this operator more closely and we
find that this expression may be re-expressed as
\begin{eqnarray}
\mathcal{V}^{-1}\sum_{{\bf n}\in \mathcal{I}} k^4_{\bf n}|\hat{\mathfrak{p}}^{{\bf n}\;l}|^2&\leq&\delta^2.
\end{eqnarray}
A useful intermediate bound is given by
\begin{eqnarray}
\mathcal{V}^{-1}\sum_{{\bf n}\in \mathcal{I}} k^4_{\bf
n} |\hat{\mathfrak{p}}^{{\bf n}\; l}|^2
&\leq&\mathcal{V}^{-1} \sum_{{\bf n}\in
\mathcal{I}}\delta^2_{{\bf n}}\leq\delta^2. \label{redelta}
\end{eqnarray}
There are a finite number of small parameters in the sum in
(\ref{redelta}), meaning a convergent sum.
This sum is bounded by $\delta^2$ and vanishes as $\delta\to 0$.  We
can just as well take every $\delta_{\bf n} \to 0$.  In this case,
$\mathbb{E}^l$ may be expanded as
\begin{equation}
\mathbb{E}^l=\prod_{{\bf n}\in \mathcal{I}} \mathbb{E}^{{\bf
n}\; l}\left[k^4_{\bf n}|\hat{\mathfrak{p}}^{{\bf n}\;l}|^2\leq
\delta^2_{\bf n}\right]. \label{eprod}
\end{equation}
Using (\ref{eprod}) in (\ref{Kdelta}), results in
\begin{align}
\mathcal{K}_1&\equiv \lim_{\delta \to 0}\frac{ \langle
\mathfrak{p}'^l, \mathfrak{h}'^l|\mathbb{E}^l \left[ \Sigma k^4_{\bf
n}|\hat{\mathfrak{p}}^{{\bf n}\; l}|^2\leq \delta ^2 \right]|
\mathfrak{p}^l, \mathfrak{h}^l\rangle} {\langle 0, 0|
\mathbb{E}^l\left[ \Sigma k^4_{\bf n}|\hat{\mathfrak{p}}^{{\bf n}\;
l}|^2\leq \delta ^2 \right]|0,0 \rangle }
\\&= \prod_{{\bf n}\in \mathcal{I}}
\lim_{\delta_{\bf n}\to 0} \frac{ \langle \mathfrak{p}'^l_{\bf n},
\mathfrak{h}'^l_{\bf n}|\mathbb{E}^{{\bf n} \;l} \left[k^4_{\bf n}
 |\hat{\mathfrak{p}}^{{\bf n}\; l}|^2\leq \delta_{\bf n}^2 \right]|
 \mathfrak{p}^l_{\bf n},
\mathfrak{h}^l_{\bf n}\rangle} {\langle 0, 0| \mathbb{E}^{{\bf n} \;l}\left[
k^4_{\bf n}
 |\hat{\mathfrak{p}}^{{\bf n}\; l}|^2 \leq \delta_{\bf n}^2 \right] |0,0 \rangle
 } \label{induct1}
 \end{align}
It is then sufficient to evaluate the reproducing kernel for some
${\bf n}$ pair ${\bf b},-{\bf b}\in \mathcal{I}$ of (\ref{ndefn}).
For each point, a resolution of unity
 may be inserted into (\ref{induct1})
to give
\begin{align}
\prod_{{\bf n}={\bf b,-b}} &\langle \mathfrak{p}'_{\bf n}, \mathfrak{h}'_{\bf n}|\mathbb{E}^{{\bf
n} \;l} \left[k^4_{\bf n}
 |\hat{\mathfrak{p}}''|^2\leq \delta_{\bf n}^2 \right]|
 \mathfrak{p}_{\bf n} ,
\mathfrak{h}_{\bf n} \rangle \nonumber\\
&=\mathcal{M}\int^{\delta_{\bf b}/k^2_{\bf b}}_{-\delta_{\bf b}/k^2_{\bf b}}
d\bar{\mathfrak{p}}'' d\mathfrak{p}'' \exp \left\{ - 4\mathcal{V}^{-1}k^4_{\bf b}\left[ \frac{|\mathfrak{p}'_{\bf b} -\mathfrak{p}'' |^2}{2} + 2i \Re (\bar{\mathfrak{p}}'' [
\mathfrak{h}'_{\bf b} -\mathfrak{h}_{\bf b}] )+
\frac{|\mathfrak{p}'' - \mathfrak{p}_{\bf b}
|^2}{2}\right] \right\},\nonumber\\
\;\;\;\quad &\times  \exp \left\{ - 4i\mathcal{V}^{-1}k^4_{\bf b}\Re ( \mathfrak{p}'_{\bf b}
\mathfrak{h}'_{\bf b}-
\mathfrak{p}_{\bf b}\mathfrak{h}_{\bf b})\right\},\label{induct2}
\end{align}
where we have suppressed all $^l$ superscripts, used $\Re (\cdot)$ to indicate the operation of extracting the real part,  and assumed a Gaussian form for the fiducial vector $|\eta^l\rangle$ in (\ref{LCS}).  Also, the finite, constant factor $\mathcal{M}$ in (\ref{induct2}) need not be determined as it cancels out when (\ref{induct2}) is plugged back into (\ref{induct1}).  Expression
(\ref{induct2}) allows us to observe the reduction
 \begin{align}
 \lim_{\delta_{\bf b}\to 0}
&\prod_{{\bf n}={\bf b,-b}} \frac{\langle\mathfrak{p}'_{\bf n},
\mathfrak{h}'_{\bf n}|\mathbb{E}^{\bf n} \left[k^4_{\bf n}
 |\hat{\mathfrak{p}}^{{\bf n}}|^2\leq \delta_{\bf n}^2 \right]|
 \mathfrak{p}_{\bf n} ,
\mathfrak{h}_{\bf n} \rangle}{\langle 0,0|\mathbb{E}^{{\bf n} }
\left[k^4_{\bf n}
 |\hat{\mathfrak{p}}^{{\bf n}}|^2\leq \delta_{\bf n}^2
 \right]|0,0\rangle}\nonumber\\
 &=\exp \left\{-
 4\mathcal{V}^{-1}k^4_{\bf
b}\left[\frac{|\mathfrak{p}'_{\bf b}  |^2}{2} +
\frac{|\mathfrak{p}_{\bf b} |^2}{2} \right]
\right\} \exp \left\{ - 4i\mathcal{V}^{-1}k^4_{\bf b}\;\Re(\mathfrak{p}'_{\bf b}
\mathfrak{h}'_{\bf b}-
\mathfrak{p}_{\bf b}\mathfrak{h}_{\bf b})\right\}\label{induct3}.
\end{align}
It then follows that
\begin{align}
\mathcal{K}&_1
[\mathfrak{p}',\mathfrak{h}';\mathfrak{p},\mathfrak{h}]\\
&=\exp \left\{ -\mathcal{V}^{-1}\sum_{{\bf n}\in
\mathcal{I}} 4k^4_{\bf n}\left[ \frac{|\mathfrak{p}'_{\bf n}
|^2}{2} + \frac{|\mathfrak{p}_{\bf n} |^2}{2}+i\Re\left(
\mathfrak{p}'_{\bf n}\mathfrak{h}'_{\bf n}-
\mathfrak{p}_{\bf n}\mathfrak{h}_{\bf n}\right)\right] \right\} .
 \end{align}

The fiducial vector representative can be written in terms of the reproducing kernel as
\begin{equation}
\Psi_0[\mathfrak{p}^l, \mathfrak{h}^l]= \mathcal{K}_1 [\mathfrak{p}^l,
\mathfrak{h}^l;0,0]= \exp \left\{ -\mathcal{V}^{-1}\sum_{{\bf n}\in \mathcal{I}} 4k^4_{\bf
n}\; \left[\frac{|\mathfrak{p}^{l}_{\bf n}|^2}{2}+
i\Re(\mathfrak{p}^{l}_{\bf n}\mathfrak{h}^{l}_{\bf n})  \right]\right\}.
\end{equation}
Writing down any member of a dense set $\Psi\in D_{\mathfrak{H}_l}$
of this Hilbert space $\mathfrak{H}_l$ leads to a vector
\begin{eqnarray}
\Psi[\mathfrak{p}^l, \mathfrak{h}^l]&=& \sum_m \beta_m \mathcal{K}_1[\mathfrak{p}^l, \mathfrak{h}^l;
\mathfrak{p}^l_m, \mathfrak{h}^l_m ]\nonumber \\
&=& \sum_m \beta_m \exp \left\{ - \mathcal{V}^{-1}\sum_{{\bf n}\in
\mathcal{I}} 4k^4_{\bf n}\left[ \frac{|\mathfrak{p}^l_{\bf n}
|^2}{2} + \frac{|\mathfrak{p}^l_{{\bf n}m} |^2}{2}+i\Re\left(
\mathfrak{p}^l_{\bf n}\mathfrak{h}^l_{\bf n}-\mathfrak{p}^l_{{\bf
n}m}
\mathfrak{h}^l_{{\bf n}m}\right)\right] \right\}\nonumber \\
&=& \Psi_0[\mathfrak{p}^l, \mathfrak{h}^l] \sum_m \beta'_m
\end{eqnarray}
which is
proportional to the fiducial vector representative in the same way
that we obtained (\ref{Psi}) earlier. The
 Hilbert space $\mathfrak{H}_l$, like $\mathfrak{H}_T$ and $\mathfrak{H}_t$, is equivalent to the space of complex numbers $\mathbb{C}$. This completes
the proof. $\Box$

\section{Appendix: Fundamental classical and quantum algebras} \label{C}

\subsection{Classical Poisson algebra}

 The following equations give the classical Poisson
bracket algebra for the tensor components of
$\{\widetilde{p}_{ab}({\bf k}), \widetilde{h}_{ab}({\bf k})\}$.

{\bf Transverse:}
\begin{eqnarray}
\left\{\bar{\widetilde{{\sf h}}}(t, {\bf k}), \widetilde{{\sf p}}(t, {\bf k}')\right\} = 2\delta^3({\bf k}-{\bf k}').\label{B1}
\end{eqnarray}

{\bf Longitudinal:}
\begin{eqnarray}
\left\{ \bar{\widetilde{\mathfrak{h}}}_1\:^t (t, {\bf k}),
\widetilde{\mathfrak{p}}_1\:^t (t, {\bf k}')\right\}&=&\left\{ \bar{\widetilde{\mathfrak{h}}}_2\:^t(t, {\bf k}), \widetilde{\mathfrak{p}}_2\:^t  (t, {\bf k}')\right\}=\frac{\delta^3({\bf k}-{\bf k}')}{4k^2}\\
\left\{ \bar{\widetilde{\mathfrak{h}}}\:^l (t, {\bf k}),
\widetilde{\mathfrak{p}}\:^l (t, {\bf k}')\right\}&=&\frac{\delta^3({\bf k}-{\bf k}')}{4k^4}
\end{eqnarray}
In the above equation and throughout the paper,
we make use of the following abbreviation for the transverse components of $\mathfrak{f}_a$:
\begin{equation}
\widetilde{\mathfrak{f}}_a=\frac{m_a \widetilde{\mathfrak{f}}^{\;t}_1 + \bar{m}_a
\widetilde{\mathfrak{f}}^{\;t}_2}{2}.
\end{equation}

{\bf Transverse-Traceless:}
\begin{eqnarray}
\left\{ \bar{\widetilde{{\bf h}}}\:^+(t, {\bf k}), \widetilde{{\bf p}}^+(t, {\bf k}')\right\}=\left\{ \bar{\widetilde{{\bf h}}}\:^-(t, {\bf k}), \widetilde{{\bf p}}^-(t, {\bf k}')\right\}=\delta^3({\bf k}-{\bf k}').\label{B4}
\end{eqnarray}
The $\widetilde{{\bf f}}^+$ and $\widetilde{{\bf f}}^-$ come from the expansion
\begin{equation}
\widetilde{{\bf f}}_{ab}=\widetilde{{\bf f}}^+m_am_b + \widetilde{{\bf f}}^- \bar{m}_a \bar{m}_b.
\label{B5}
\end{equation}

\subsection{Quantum commutation relations}

The fundamental commutation relations may then be determined for each set of operator components,
using the lattice
prescription of Section 3.1.

{\bf Transverse:}
\begin{eqnarray}
\left[\bar{\hat{{\sf h}}}^{\bf m}, \hat{{\sf p}}^{\bf n}\right] =
2i\mathcal{V}\delta^{{\bf m,n}}.\label{B1}
\end{eqnarray}

{\bf Longitudinal:}
\begin{eqnarray}
\left[ \bar{\hat{\mathfrak{h}}}_1\:^{{\bf m}t} ,
\hat{\mathfrak{p}}_1\:^{{\bf n}t} \right]&=&\left[ \bar{\hat{\mathfrak{h}}}_2\:^{{\bf m}t}, \hat{\mathfrak{p}}_2\:^{{\bf n}t}  \right]
=\frac{i\mathcal{V}\delta^{{\bf m,n}}}{4k^2_{\bf m}}\\
\left[ \bar{\hat{\mathfrak{h}}}\:^{{\bf m}l} ,
\hat{\mathfrak{p}}\:^{{\bf n}l}
\right]&=&\frac{i\mathcal{V}\delta^{{\bf m,n}}}{4k^4_{\bf n}}
\end{eqnarray}

{\bf Transverse-Traceless:}
\begin{eqnarray}
\left[ \bar{\hat{{\bf h}}}\:^{{\bf m}+}, \hat{{\bf p}}^{{\bf
n}+}\right]=\left[ \bar{\hat{{\bf h}}}\:^{{\bf m}-}, \hat{{\bf
p}}^{{\bf m}-}\right]=i\mathcal{V}\delta^{{\bf m,n}}.\label{B4}
\end{eqnarray}
Again, the $\hat{{\bf f}}^+$ and $\hat{{\bf f}}^-$ come from the expansion
\begin{equation}
\hat{{\bf f}}_{ab}=\hat{{\bf f}}^+m_am_b + \hat{{\bf f}}^- \bar{m}_a
\bar{m}_b \label{B5}
\end{equation}

\section*{Acknowledgements}

The authors would like to thank Abhay Ashtekar, for helpful comments
and questions surrounding the $\delta \to 0$ limit; J. Scott Little
for help fruitful useful discussions; and Sergei Shabanov on the
application of the Projection Operator to field theories.

\end{document}